# Influence of Ru-doping on structure, defect chemistry, magnetic interaction and carrier motion of the $La_{1-x}Na_xMnO_{3+\delta}$ manganite

Lorenzo Malavasi[a,*], Maria Cristina Mozzati[b], Cristina Tealdi[a], Carlo Bruno Azzoni[b], and Giorgio Flor[a]

[a] *Dipartimento di Chimica Fisica "M. Rolla", INSTM, IENI/CNR Unità di Pavia of Università di Pavia, V.le Taramelli 16, I-27100, Pavia, Italy.*

**E-mail: lorenzo.malavasi@unipv.it*

[b]*CNISM and Dipartimento di Fisica "A. Volta", Università di Pavia, Via Bassi 6, I-27100, Pavia, Italy.*

## Abstract

In this work we report a structural, electrical and magnetic characterization of the $La_{1-x}Na_xMn_{1-y}Ru_yO_{3+\delta}$ (LNMRO) system with $x$ = 0.05, 0.15 and $y$ = 0, 0.05, 0.15, also comprising an investigation of the role of the oxygen content on the related redox properties. The experimental investigation has been realized with the aid of X-ray powder diffraction, electron microprobe analysis, thermogravimetry, electrical resistivity and magnetization measurements, and electron paramagnetic resonance. We demonstrate that the effect of ruthenium doping on the studied LNMRO compounds is not only directly related to the Ru/Mn substitution and to the Ru oxidation state but also indirectly connected to the oxygen content in the sample. Our data show that ruthenium addition can "improve" electrical and magnetic properties of non-optimally (low) cation doped manganites, causing an increase of the $T_C$ value and the insurgence of MR effect, as observed for the $x$ = 0.05 and $y$ = 0.05 sample (MR ≅ 60% at 7 T and at about 260 K).

.

*Keywords*: Manganite, Perovskites, Ruthenium, Defects, Magnetoresistance, Na-doped manganite

## 1. Introduction

Strong interest has been triggered, in recent years, by the possibility of tuning the structural, magnetic and electrical properties of mixed-valent manganite perovskites by cation doping on the Mn-site, so that new peculiar features were discovered as the appearance of a metallic-like state and of ferromagnetism in insulating and antiferromagnetic manganites when doped with chromium [1,2], cobalt [3], nickel [3], or ruthenium [4-8]. This effect was connected to the "collapse" of the charge and orbital ordering of $Mn^{3+}$ and $Mn^{4+}$ ions.

It was shown by Vertuyen *et al.* [8] that $Ni^{2+}$ substitution in $LaMnO_3$ induces double exchange $Mn^{3+}/Mn^{4+}$ ferromagnetic (FM) interactions competing with the $Mn^{3+}/Mn^{3+}$ antiferromagnetic (AF) super-exchange interactions: $Ni^{2+}$ takes place itself directly in the exchange interactions, leading to a stabilization of the FM ground state. Other "new" phenomena have been found in Cr-doped manganites: for the $Pr_{0.5}Ca_{0.5}Mn_{0.99}Cr_{0.01}O_3$ compound Raveau *et al.* [2] reported about a "spectacular" resistivity increase in a 4 T magnetic field: also in this case the main reason was accounted for by the partial destruction of the charge ordering and by phase separation.

Among the several possible dopants, one of the most studied is the ruthenium. Main reasons for this can be summarized as follows: i) a large Ru amount can be introduced within the perovskite network keeping it single-phase; ii) Ru can show different oxidation states ($Ru^{3+}$, $Ru^{4+}$ and $Ru^{5+}$); iii) the $Ru^{3+}$ and $Ru^{4+}$ low-spin configuration make this dopant suitable for FM coupling with the Mn-ions thus preserving and participating to the double-exchange mechanism.

Ru substitution can induce, in a very large composition range [4], metallic-behavior and FM in AF materials irrespective to the exact nature of the AF order. However, up to now, several "anomalies" in the metallic (M) and FM state of Ru-doped manganites are far from being understood; among them we can cite the small values of the experimental magnetic moment (even in high magnetic fields) with respected to the calculated ones and the "double bump" in the resistivity curves $\rho(T)$. Some argumentation are given in the current literature in order to explain all

- or at least some - of the above cited features. Raveau *et al.* [4] argue that Ru substituted for Mn is in the +5 valence state in agreement with other experimental results [9], and consequently the doping will increase the $Mn^{3+}$ content according to:

$$2Mn^{4+} \rightarrow Ru^{5+} + Mn^{3+} \qquad (1)$$

or, if more correctly written according to a Kroeger-Vink notation:

$$2Mn_{Mn}^{x} + 2RuO_2 \Leftrightarrow 2Ru_{Mn}^{\bullet} + 2e^- + Mn_2O_3 + 0.5O_2 \qquad (2a)$$

$$Mn_{Mn}^{x} + e^- \Leftrightarrow Mn_{Mn}^{-} \qquad (2b)$$

$$Mn_{Mn}^{\bullet} + e^- \Leftrightarrow Mn_{Mn}^{x} \qquad (2c)$$

Note that these equations do not imply the formation of any point defect (except for the substitutional defect), but only the occurrence of a redox reaction between Mn and Ru ions. In particular, equation 2b indicates the possible reduction of $Mn^{3+}$ to $Mn^{2+}$ while equation 2c is relative to the reduction of $Mn^{4+}$ ions to $Mn^{3+}$.

The model suggested by Raveau is able to explain the FM interaction between the *newly* generated $Mn^{3+}$ and the $Mn^{4+}$ when the starting material is AF and constituted only by tetravalent ions like $CaMnO_3$. In addition, the same Authors indicate the possible valence fluctuation

$$Mn^{4+} + Ru^{4+} \rightarrow Ru^{5+} + Mn^{3+} \qquad (3)$$

as a source of FM interaction between Ru and Mn ions and by considering a possible hybridization between the $Ru^{5+}$ and $Ru^{4+}$ (electronic configurations $t_{2g}^3$ and $t_{2g}^4$, respectively) empty $e_g$ orbitals and the $Mn^{3+}$ orbitals thus directly participating in a FM exchange.

Other Authors [8,10,11] suggest that the metal-insulator (M-I) transitions and the magnetoresistance (MR) behavior found in Ca-doped $LaMnO_3$ can be understood in the framework of phase separation and coexistence of two FM and M phases: one $Ru^{4+}$ rich with FM coupling between Ru and Mn ions, and another (with FM transition at lower temperature and poor conducting in the magnetic phase) characterized by the presence of $Ru^{3+}$ ions AF-coupled with neighboring Mn ions. This overall picture indicates that, up to now, a clear explanation of the *real* effect caused by Ru-ion when entering the perovskite structure is still lacking.

We already carried out deep studies on Na-doped manganites as a function of composition ($x$), oxygen content ($\delta$) and preparation route [12-17]. So, in order to understand and deepen the role of Ru substituted on Mn site, we started a thorough study on the $La_{1-x}Na_xMn_{1-y}Ru_yO_{3+\delta}$ system (LNMRO) with optimally cation doping, *i.e.* with average valence state of Mn ~ 3.30, and with lower cation doping.

We also stress that our previous X-ray absorption spectroscopy (XAS) investigation on this system [18] demonstrated that the ruthenium is present, in hole doped mixed-valence manganites of the $La_{1-x}Na_xMnO_{3+\delta}$ system, with a mean valence state ≤ +4. XAS measurements clearly indicated that a simple Mn reduction (as equations 2a, b and c) is not the only mechanism involved in charge compensation when Ru is introduced in the manganite lattice, also according to the absence of a Mn valence state lower than +3. We then suggested that cation vacancies compensation through:

$$2Mn_{Mn}^{X} + 2RuO_2 \Leftrightarrow 2Ru_{Mn}^{\bullet} + O_O^X + \frac{1}{3}V_{Mn}''' + \frac{1}{3}V_{La}''' + Mn_2O_3 \qquad (4)$$

acts, in particular at a relatively low intrinsic hole doping ($x$ = 0.05), to keep the system neutral while at higher holes concentration a more direct electron exchange between $Mn^{3+}/Mn^{4+}$ and $Ru^{3+}/Ru^{4+}$ couples is present. In both cases, the mechanisms involved assure to avoid a too strong manganese reduction. Moreover, an efficient redox system is present in the $x$ = 0.15 samples which allows to keep the average Mn valence state nearly constant even though lower with respect to the Ru-free compound.

In a previous Communication [19] we already evidenced how the currently suggested charge compensation mechanisms for the Ru-doping fail to account for all the properties of these samples. In this extended work we will report a more complete structural, electrical and magnetic characterization on the LNMRO system, also comprising an investigation of the role of the oxygen content on the related redox properties. The experimental investigation has been realized with the aid of several techniques such as X-ray powder diffraction (XRPD), electron microprobe analysis (EMPA), thermogravimetry (TG), electrical resistivity and magnetization measurements, and electron paramagnetic resonance (EPR).

## 2. Experimental

Samples of the $La_{1-x}Na_xMn_{1-y}Ru_yO_{3+\delta}$ (LNMRO) system were prepared with $x$ = 0.05 and 0.15 and $y$ = 0, 0.05 and 0.15 by solid state reaction starting from $La_2O_3$ (Aldrich, 99.999%), $Mn_2O_3$ (Aldrich, 99.999%), $Na_2CO_3$ (Aldrich, 99.99%) and $Ru_2O_5$ (Aldrich 99.99%). Pellets were prepared from the thoroughly mixed powders and allowed to react at 1233 K for at least 90 hours, during which were re-ground and re-pelletised at least twice. Low reaction temperature and long firing times have been chosen in order to assure no Na evaporation during the synthesis [12].

After synthesis, two batches of the as-prepared samples were annealed for 72 h at 1173 K in oxygen and in argon, respectively, and quenched to room temperature in an ice-water mixture.

XRPD patterns were acquired on a Bruker "D8 Advance" diffractometer equipped with a Cu anticathode, graphite monochromator on the diffracted beam, and proportional detector. Measurements were carried out in the angular range from 10° to 120° with a step size of 0.02° and a counting time of 10 s per step. Diffraction patterns were refined by means of Rietveld method with the FULLPROF software [20].

EMPA measurements were carried out using an ARL SEMQ scanning electron microscope, performing at least 10 measurements in different regions of each sample. According to EMPA and XRPD data, the above synthetic procedure gave single-phase materials; in addition each sample was found to be highly homogeneous in the chemical composition, which was met in good agreement with the nominal one.

In TG measurements, performed to determine the weight variation of the samples when exposed to different atmospheres, the samples underwent the following treatments:

i) a 10 K/min ramp in oxygen;

ii) an isothermal step at 1073 K in oxygen for 360 minutes;

iii) a 10 K/min ramp in oxygen down to 423 K;

iv) a 10 K/min ramp in argon up to 1173 K .

During the steps i) and ii) the samples are allowed to equilibrate with the oxygen atmosphere and increase their oxygen content of a $\delta$ amount; in addition, by reducing the temperature during step iii), the maximum $\delta$-value, relative to these experimental conditions, is achieved. Finally, the step iv) in argon induces the system to equilibrate within the new conditions, thus leading to a loss of oxygen.

Absolute values of oxygen contents were determined by means of chemical analysis on some selected samples through the commonly used titration with potassium permanganate [21] and through previous thermo-gravimetric results on the Ru-undoped samples. We remark, however, that the presence of two redox couples makes the definition of the oxygen content a quite hard task. So we are aware of an average relative error of 5-10% on the $\delta$-values presented. Let us note that the observed weight variations are related to the bulk and not to the surface, because of the extremely low value of the surface area.

Static magnetization was measured at different applied fields in the temperature range 2-350 K with a SQUID magnetometer (Quantum Design). Field dependence of magnetization was also investigated at 2K for field ranging between 0 and 7 T.

Resistivity and MR measurements were carried out between 320 and 10 K at 0, 1 and 7 T magnetic fields with DC-four electrodes method by means of a specific probe directly inside the magnetometer. Complementary EPR measurements were performed in X band (~ 9.5 GHz) to study the temperature dependence in the range 300 - 470 K.

## 3. Results

### *1. X-ray powder diffraction (XRPD)*

XRPD patterns were collected on all the LNMRO samples for both oxygen and argon thermal treatments. As a selected example, Figure 1 reports the Rietveld refined pattern of the $La_{0.95}Na_{0.05}Mn_{0.95}Ru_{0.05}O_{3+\delta}$ sample annealed in oxygen, from now named N5R5O. The crystal structure is rhombohedral and belongs to the *R*-3*c* space group (no. 67). Lattice constants, cell volumes, crystal structure and oxygen content for this sample and for all the other ones are listed in Table 1.

All the samples treated in oxygen, irrespective to the Na or Ru doping, posses a rhombohedral structure. Looking at the $x$ = 0.05 series it can be noticed that the $a$($c$) parameter has a minimum(maximum) in correspondence of the Ru = 0.05 sample. On the opposite, for the $x$ = 0.15 series, an overall expansion of both parameters occurs by increasing the Ru-doping.

The treatment in argon causes, for $x$=0.05, the material to adopt an orthorhombic structure (s.g. *Pbnm*) while, for $x$=0.15, only the sample with $y$ = 0.15 becomes orthorhombic, while the others remain rhombohedral. Regarding the $x$ = 0.05 series, it can be noticed that the orthorhombic distortion is more severe (O′) for the undoped and $y$=0.15 samples ($b/a<\sqrt{2}$), while the common O-distortion ($b/a>\sqrt{2}$) pertains to the $y$=0.05 compound. Let us recall that in the O′–phase a cooperative and static Jahn-Teller distortion, resulting from the $e_g$ ($x^2$-$y^2$) orbital ordering, leads to the expansion of the unit cell in the basal plane.

In Table I the volume ($V$) for all the samples is reported. For both the series, the argon treatment causes a general $V$-increase as a consequence of the reduction of both Mn and Ru ions [18]. In more detail, for the samples with $x$=0.05 annealed in pure oxygen only a slight variation of $V$ is found as the Ru-content ($y$) increases while a first decrease followed by an increase is observed for the $x$=0.05 samples annealed in argon. A more "easy" $V$-trend is found for the $x$=0.15 samples since for both the annealing treatments $V$ increases by increasing $y$.

## *2 Resistivity and magnetoresistivity*

Referring to the $x$ = 0.05 samples annealed in oxygen, the Ru-free sample (N5O) is semiconducting (S) while the Ru addition effectively reduces the resistivity of the samples (also inducing S-M transition), as already shown [19]. By applying a magnetic field, no significant $\rho$-variation was found for the N5O sample while an appreciable effect was observed for the N5R5O and N5R15O samples. The $\rho(T)$ curves at 0, 1 and 7 T and the correspondent MR($T$) curves for these two samples are plotted in Figures 2a and 2b, respectively. The MR values were calculated according to:

$$MR(\%) = \frac{R(H) - R(0)}{R(0)} \bullet 100 . \qquad (5)$$

For the N5R5O sample the magnetic field strongly reduces the resistivity, with an evident MR effect at a $T$ value corresponding to the S-M transition at higher temperature, while no clear MR behavior is detectable in connection with the wide bump in the $\rho(T)$ curve. For N5R15O the application of a magnetic field allows to detect the two transitions seen as shoulder in the $\rho(T)$ curves. All the argon-annealed samples with x = 0.05 are insulators with very high $\rho$-values and with negligible MR effects. We just stress that also in this case the Ru-doped samples have $\rho$-values lower with respect to the Ru-free ones.

Let us now consider the $x$ = 0.15 samples. For all the oxygen annealed ones a S-M transition is present, as shown in the Figures 3a, 3b, 4a. From these curves we can deduce, as a common trend, a decrease of the $\rho(T)$ and transition temperature ($T_{SM}$) by increasing $y$. The argon annealed samples of $x$=0.15 series generally show a $T_{SM}$ decrease or a semiconducting behaviour. Nevertheless, an interesting MR effect is observed for N15R15Ar, which also shows an increase of $T_C$ (see later) upon reducing condition annealing. For N15R15Ar, the MR, plotted in Figure 4b together with the related ln$\rho(T)$ curves at 0, 1 and 7 T, shows a clear peak around 200 K, whose position slightly moves towards higher $T$ by increasing the magnetic field. Differently from all the other samples a net decrease of the MR response is observed before and after the peak.

### *3 Magnetic measurements*

Figure 5a reports the zero field cooling (ZFC) and field cooling (FC) molar susceptibility $\chi_{mol}$ *vs.* $T$ curves at 100 Oe for the samples with $x$=0.05 annealed in oxygen. All the samples show clear transitions from a paramagnetic (P) to a ferromagnetic (F) state. The Curie temperatures ($T_C$), taken at the inflection points on the FC curves, are reported in Table 1: the lowest $T_C$ corresponds to the Ru-undoped sample. Significant deviation between the FC and ZFC curves occurs thus indicating the presence of small magnetic domains possibly connected with magnetic un-homogeneity [12] related to the large $\delta$-values and, consequently, to the presence of a large amount of cation vacancies. Indeed, the N5R5 sample shows the best characteristics, in spite of the lowest Mn average valence according to the XAS data [18].

In Figure 5b the analogous curves but after argon annealing are reported. For the Ru-undoped sample $T_C$ is practically unchanged, while for the other two samples $T_C$ reduces of ~70 K (N5R5) and of ~45 K (N5R15), respectively. The decrease of $T_C$ for this series with respect to the oxygen annealed one has to be ascribed to a general decrease of Mn average valence state, as indicated by XAS data. Besides, the orthorhombic structure and a generally higher cell volume, correspondent to the presence of higher amount of $Ru^{3+}$, can hinder the magnetic interactions with respect the correspondent oxygen annealed samples. For both oxygen and argon annealed samples, however, the sharpest PF transition with the highest $T_C$ values occurs for the N5R5 composition.

The $x$=0.15 series is characterized by higher $T_C$ values and sharpest magnetic transitions with respect to the $x$=0.05 series. In Figure 6a the FC and ZFC curves at 100 Oe for the samples with $x$ = 0.15 annealed in oxygen are reported. It can be noticed a progressive $T_C$ reduction by increasing the Ru-doping together with a widening of the PF transition and increased difference between the FC and ZFC curves. For these samples, the presence of $Ru^{4+}$ (according to XAS results) is indeed mainly related to a progressive increase of the oxygen overstoichiometry, so giving rise to progressive magnetic un-homogeneity in the samples. The argon treatment causes a $T_C$ reduction for the undoped and the N15R5 sample and, on the opposite, an increase for the N15R15 one (Figure 6b). The sharper PF transition and the higher $T_C$ value of this sample with respect to the analogous oxygen annealed one suggests again the important role of the cation vacancies on the magnetic properties of these compounds ($\delta$ is reduced of ~30% with respect the N15R15O sample).

For all the considered samples, saturation of the magnetisation was reached at 2 K by applying magnetic field up to 7 T. The magnetic moment value per unit formula obtained from magnetization saturation are in good agreement (with an error less than 10%) with the expected ones from the stoichiometric formula, considering $S$=0 for $Ru^{4+}$ ions. A further improvement of the agreement is obtained considering the vacancies formation mainly on B site, as suggested by neutron diffraction results found in the Na doped manganites [22].

To investigate the magnetic (and chemical) homogeneity of the samples and to get information about the character of ruthenium/manganese substitution EPR measurements were performed. The Ru-undoped samples showed the typical lanthanum manganite spectrum, already observed and described in [12,23,24]. For all the investigated samples a signal with $g$ factor ≅ 2 is present at room temperature, as due to $Mn^{4+}$ or $Mn^{3+}$-$Mn^{4+}$ Zener pairs. In many cases, a second component is detectable with effective $g$ factor, $g_{eff}$, higher than 2 that moves towards $g$ ≅ 2 by increasing the temperature. This suggests the presence of sample regions with different long range magnetic interactions, these regions merging more or less rapidly accordingly to the sharpness of the PF transition. EPR signals related to ruthenium are not clearly observable in all the samples. For the N5 samples a structured signal with $g>2$ is present while for the N15 samples this signal is present only if argon annealed. These data can agree with the XAS results [18] by considering that $Ru^{4+}$ is not EPR active and ascribing the structured signal to $Ru^{3+}$.

Starting from the data previously obtained for the Ru-undoped compounds, an evaluation of the $Mn^{4+}$ content of the present samples has been drawn by comparing the EPR intensities (areas) at 470 K, *i.e.* in the paramagnetic phase, and taking into account the dependence of the EPR intensity on the $T_C$ values. A very good agreement can be found with XAS results when considering low Ru content (≤5%), whereas some discrepancies are revealed for the samples with the highest Ru content where this procedure ended up with an estimation of the $Mn^{4+}$ amount lower than 30-35% with respect to the one determined from XAS data. The reason for these differences can be related to the different temperatures considered for XAS (*RT*) and EPR (470K) data, particularly if the internal redox reaction:

$$Mn^{4+} + Ru^{3+} \rightarrow Mn^{3+} + Ru^{4+} \qquad (6)$$

may occur already at 470 K, thus affecting the EPR intensity. However, this last point will be object of future research in order to define any possible valence fluctuation at high-*T* or other effects that may lead to this discrepancy only for the Ru=0.15 samples.

# 4 – Discussion and Conclusion

In this paper we presented the results of an investigation of the structural, electrical and magnetic properties of the $La_{1-x}Na_xMn_{1-y}Ru_yO_{3+\delta}$ system (LNMRO) with optimally cation doping, ($x$ =0.15) and with lower cation doping ($x$=0.05).

Several parameters affect the properties of this system that can be considered, then, a *complex* system. Among them:

i) *Na-doping*, which broads the conduction band by creating holes;

ii) *Ru-doping*, which on one hand contributes positively to the DE-FM exchange thanks to the availability of $e_g$ orbitals, in particular when it is diluted in the B-sublattice, while for higher concentration we may expect SE (super exchange)-AF coupling between $Mn^{3+}$ and $Ru^{n+}$ ions; on the other hand, the Ru-doping involves the Mn-reduction and/or the oxygen content increase which in turn means an increase of the cation vacancy concentration; moreover, a broadening of the conduction band may be expected due to the more expanded nature of 4$d$ orbitals of Ru;

iii) *Oxygen content*, which, as already recognized for Ru-free materials, is a key factor in manganites but here plays an additional role being involved in partial or total Ru-compensation; in addition, holes and electrons created by oxygen over- or under-stoichiometry showed to affect in a not trivial way the redox couples present in the LNMRO series;

iv) $e_g$ *occupation*, which may play a fundamental role in determining the possible hole hopping between Zener ions; all the Ru ions, irrespective to their oxidation state, have unoccupied $e_g$ orbitals and can be regarded as $Mn^{4+}$ ions. Nevertheless, possible AF interaction between $Mn^{4+}$ and $Ru^{n+}$ ions may become progressively significant for high Ru-dopings and high Mn valence states.

However, the points listed are not exhaustive since, for example, the role of ion mismatch on both the A and B sites was not taken into account. We will try, in the following discussion, to highlight which are the most important effects for the different series of samples studied.

Let us start with Na=0.05 samples annealed in pure oxygen. As shown in Figure 7, the maximum $T_C$ (251 K) was found for intermediate Ru-doping (0.05) while Ru-free sample has a $T_C$ lower of more than 100 K and Ru=0.15 sample of about 40 K, notwithstanding these last two samples have a Mn average valence higher than that of the Ru=0.05 compound. Major role is then played here by cation vacancies and Ru-doping. N5R5O has fewer cation vacancies with respect to both N5O and N5R15O compounds since part of the charge compensation due to the Ru-doping is accomplished by partial internal Mn-reduction. Moreover, with respect to the N5O sample, it contains Ru ions which contribute effectively in the enhancement of $T_C$. Differently, the N5R15O sample has more Ru but also more cation vacancies: the $T_C$ difference between N5R5O and N5R15O samples is more or less the one expected for a difference of about 0.5% in vacancies concentration [25]. In this case, the effect of the interruption of interaction path caused by the vacancies reflects also in the insurgence of a semiconducting-like state for the N5R15O samples with respect to the N5R5O.

Let us note the presence of a double $\rho$-transition in both these two samples. Previous Authors correlated it to a magnetic phase separation. The application of a magnetic field shifts at higher values the position of the first transition, namely the one at higher $T$, as usually found for S-M transitions, while it does not move the position of the second one. This suggests that this second transition may not be joined to a magnetic transition. We propose that the origin of this second "bump" in the $\rho(T)$ is actually related to the point defects connected to the compensation mechanism induced by the Ru-doping. The effect caused be cation vacancies on the shape of $\rho(T)$ curves was already pointed out for LCMO and LNMO Ru-undoped systems [12,26] where disorder-induced localization phenomena where the source of the resistivity increase and of the appearance of shoulders in the curves.

The Na=0.15 series shows nicely the direct role of charge compensation through the mechanism (4) and partial internal redox reaction between Mn and Ru ions. The progressive reduction of Mn valence state and increase of the oxygen content cause the trend in the Curie temperatures observed in Figure 7. Ru-doping causes the appearance of a bump in the $\rho(T)$ curve of the N15R5O sample and of a semiconducting-like regime for the higher Ru content. It is clear that for optimally doping, where all the hole doping is created through the extrinsic cation doping (Na), the Ru-addition deteriorates the system properties. This happens because more and more cation vacancies are introduced along with the Ru-doping. However, also the Ru itself does not "help" the system, in this case, since the hole doping is already optimal and, clearly, achieving the same amount of Zener ions with Mn ions only or with two different kinds of ions is not the same, thus suggesting that some size effect caused by the strong mismatch between Ru and Mn ions plays a role.

A final comment on the argon annealing treatments. These were mainly performed in order to study the interplay between the redox couples present and the oxygen content variation. Ru seems to play a sort of *buffer* role thus preserving the system from a too strong reduction of Mn ions. This is more evident for the Na=0.15 samples where the average Mn valence state between oxidizing or reducing thermal treatments is practically unchanged, while the valence of Ru ions is strongly shifted towards lower values. This is also the reason for the "unexpected" increase of $T_C$ for the N15R15 sample after the annealing in the argon environment, where, as usual for manganites, we would have expected a $T_C$ reduction. In this case, also an interesting MR effect is observed.

In the literature, together with the anomalous presence of two transitions in the $\rho(T)$ curves, also the magnetic moment value per unit formula was found to be significantly smaller that the expected one in CO-AFM manganites [4]. As we showed in the Results section dealing with the magnetic data, we found a good agreement between the expected and calculated magnetic moments by taking into account the exact nature of the samples, i.e. the cation stoichiometry, oxygen content and Ru valence state. So, we may suggest that discrepancy between calculated and experimental data may come from an incomplete analysis of all the samples features which, we recognize, may be a though work for these complex systems.

To conclude, the effect of ruthenium doping on the structural, transport and magnetic properties of the $La_{1-x}Na_xMn_{1-y}Ru_yO_{3+\delta}$ compounds is not only directly related to the Ru/Mn substitution and to the Ru oxidation state but also indirectly connected to the oxygen content in the sample. Our data show that ruthenium addition can "improve" electrical and magnetic properties of non-optimally (low) cation doped semiconducting manganites, causing an increase of the $T_C$ value and the insurgence of MR effect, as observed for the N5R5O sample (MR ≅ 60% at 7T and at about 260K).

The present study has shown that in order to characterize these *complex* systems and extract meaningful physical information the full "chemical state" of all the samples, defined by the interconnected parameters cation composition, oxygen content, ions valence state and spin state, must be known. Our study may be a starting point in order to suggest a proper way to get a full understanding of the effect of transition metal ions doping on the B-site of manganites.

## Acknowledgments

Dr. Simona Bigi is gratefully acknowledged for having performed EMPA analysis. The Department of earth science of Modena University and CNR of Modena are acknowledged for allowing SEM use. E. Jarosewich has kindly supplied the Lanthanum standard for EMPA measurements. This work was founded by the Italian Ministry of the University and Research through the PRIN 2004 project.

## References

[1] Cabeza O.; Long M.; Sevevac C.; Bari M.A.; Murihead C.M.; Francesconi M.G.; Greaves C. *J. Phys.: Condens. Matter* **1999**, *11,* 2569.

[2] Raveau B.; Maignan A.; Mahendiran R.; Khomskii D.; Martin C.; Hébert S.; Hervieu M.; Frésard R. *J. Phys. Chem. Solids* **2002**, *63,* 901.

[3] Hébert S.; Martin C.; Maignan A.; Retoux R.; Hervieu M.; Nguyen N.; Raveau B. *Phys. Rev. B* **2002**, *65*, art. 104420.

[4] Raveau B.; Maignan A.; Martin C.; Hervieu M. *J. Superconductivity* **2001**, *14*, 217.

[5] Singh B.; Sahu R.K.; Manoharan S.S. *J. Magn. Magn. Mater.* **2004**, *270*, 358.

[6] Autret C.; Martin C.; Maignan A.; Hervieu M.; Raveau B.; Andre G.; Bouree F.; Kurbakov A.; Trounov V. *J. Magn. Magn. Mater.* **2002**, *241*, 303.

[7] Sahu R.K.; Manoharan S.S. *J. Appl. Phys.* **2002**, *91*, 7517.

[8] Vertruyen B.; Flahaut D.; Hébert S.; Maignan A.; Martin C.; Hervieu M.; Raveau B. *J. Magn. Magn. Mater.* **2004**, *280*, 75.

[9] Battle P.D.; Jones C.W. *J. Solid State Chem.* **1989**, *78*, 281.

[10] Seetha Lakshmi L.; Sridharan V.; Natarajan D.V.; Chandra S.; Sankara Sastry V.; Radhakrishnan T.S.; Pandian P.; Justine Joseyphus R.; Narayanasamy A. *J. Magn. Magn. Mater.* **2003**, *257*, 195.

[11] Seetha Lakshmi L.; Sridharan V.; Natarajan D.V.; Rawat R.; Chandra S.; Sankara Sastry V.; Radhakrishnan T.S. *J. Magn. Magn. Mater.* **2004**, *279*, 41.

[12] Malavasi L.; Mozzati M.C.; Ghigna P.; Azzoni C.B.; Flor G. *J. Phys. Chem. B* **2003**, *107*, 2500.

[13] Malavasi L.; Mozzati M.C.; Polizzi S.; Azzoni C.B.; Flor G. *Chem. Mater.* **2003**, *15,* 5036.

[14] Malavasi L.; Mozzati M.C.; Ghigna P.; Chiodelli G.; Azzoni C.B; Flor G. *Recent Res. Devel. Physics* **2003**, *4*, 545.

[15] Bondino F.; Platè M.; Zangrando M.; Zacchigna M:; Cocco D.; Comin A.; Alessandri I.; Malavasi L.; Parmigiani F. *J. Phys. Chem. B*. **2004**, *108*, 4018.

[16] Malavasi L.; Mozzati M.C.; Alessandri I.; Depero L.E.; Azzoni C.B.; Flor G. *J. Phys. Chem. B* **2004**, *108*, 13643.

[17] Ghigna P.; Carollo A.; Flor G.; Malavasi L.; Subias Peruga G. *J. Phys. Chem. B* **2005**, 109, 4365.

[18] Malavasi L.; Mozzati M.C.; Di Tullio E.; Tealdi C.; Flor G. *Phys. Rev. B* **2005**, in press.

[19] Malavasi L.; Mozzati M.C.; Tealdi C.; Pascarelli M.R.; Azzoni C.B.; Flor G. *Chem. Commun.* **2004**, *12*, 1408.

[20] Rodriguez-Carvajal J. *Physica B* **1993**, *192*, 55.

[21] Yang J.; Song W.H.; Ma Y.Q.; Zhang R.L.; Sun Y.P. J. *Mag. Mag. Mat.* **2004**, *285*, 417.

[22] Malavasi L. ; Ritter C.; Mozzati M. C.; Tealdi C.; Islam M. S.; Azzoni C. B.; Flor G. *J. Solid State Chem*. **2005**, in press.

[23] Oseroff S.B.; Torikachvili M.; Singley J.; Ali S.; Cheong S-W.; Schultz S. *Phys. Rev. B* **1996**, *53,* 6521.

[24] Causa M.T.; Tovar M.; Caneiro A.; Prado F.; Ibañez G.; Ramos C.A.; Butera A.; Alascio B.; Obradors X.; Piñol S.; Rivadulla F.; Vázquez-Vázquez C.; López-Quintela M.A.; Rivas J.; Tokura Y.; Oseroff S.B. *Phys. Rev. B* **1998**, *58,* 3233.

[25] Vergara J.; Ortega-Hertogs R.J.; Madruga V.; Sapiña F.; El-Fadii Z.; Martinez E.; Beltrán A.; Rao K.V. *Phys. Rev. B* **1999**, *60*, 1127.

[26] Malavasi L.; Mozzati M.C.; Azzoni C.B.; Chiodelli G.; Flor G. *Solid State Commun.* **2002**, *123*, 321.

## Figure captions

**Fig. 1.** Rietveld refined X-ray pattern of the $La_{0.95}Na_{0.05}Mn_{0.95}Ru_{0.05}O_{3+\delta}$ (N5R5O) sample annealed in pure oxygen. In the inset the region around the most intense peak is magnified. The asterisks correspond to the reflections coming from the sample holder.

**Fig. 2.** Resistivity curves for the N5R5O (a) and N5R15O (b) samples measured at null applied magnetic field (0 T, solid line) and at 1 (dotted-line) and 7 T (dash-dotted-line). MR curves at 1 T (empty circles) and 7 T (empty triangles-up) are also plotted (see scale on the right y-axis).

**Fig. 3.** Resistivity curves for the N15O (a) and N15R5O (b) samples measured at null applied magnetic field (0 T, solid line) and at 1 (dotted-line) and 7 T (dash-dotted-line). MR curves at 1 T (empty circles) and 7 T (empty triangles-up) are also plotted (see scale on the right y-axis).

**Fig. 4.** Resistivity curves for the N15R15O (a) and N15R15Ar (b) samples measured at null applied magnetic field (0 T, solid line) and at 1 (dotted-line) and 7 T (dash-dotted-line). MR curves at 1 T (empty circles) and 7 T (empty triangles-up) are also plotted (see scale on the right y-axis).

**Fig. 5.** ZFC and FC curves for oxygen (a) and argon (b) annealed $La_{0.95}Na_{0.05}Mn_{1-y}Ru_{y}O_{3+\delta}$ samples with $y$=0 (circles), 0.05 (squares) and 0.15 (triangles).

**Fig. 6.** ZFC and FC curves for oxygen (a) and argon (b) annealed $La_{0.85}Na_{0.15}Mn_{1-y}Ru_{y}O_{3+\delta}$ samples with $y$=0 (circles), 0.05 (squares) and 0.15 (triangles).

**Fig. 7.** Curie temperatures ($T_C$) as a function of Ru-doping for all the samples considered. See legend for details.

## Table caption

**Table 1.** Sample formula, annealing treatment, crystal structure (R=rhombohedral; O=orthorhombic), lattice parameters ($a$, $b$ and $c$), cell volumes ($V$), oxygen content ($\delta$), and Curie temperatures ($T_c$), for the samples considered in the paper.

## Table 1

| Sample | Treatment | Structure | ***a*** (Å) | ***b*** (Å) | ***c*** (Å) | ***V*** (Å$^3$) | ***δ*** | $T_c$ (K) |
|---|---|---|---|---|---|---|---|---|
| $La_{0.95}Na_{0.05}MnO_3$ | Oxygen | R | 5.5241(2) | 5.5241(2) | 13.3417(2) | 58.75(1) | 0.078 | 117 |
| $La_{0.95}Na_{0.05}Mn_{0.95}Ru_{0.05}O_3$ | Oxygen | R | 5.5212(2) | 5.5212(2) | 13.3501(3) | 58.744(1) | 0.04 | 251 |
| $La_{0.95}Na_{0.05}Mn_{0.85}Ru_{0.15}O_3$ | Oxygen | R | 5.5263(2) | 5.5263(2) | 13.3455(2) | 58.824(2) | 0.058 | 212 |
| $La_{0.95}Na_{0.05}MnO_3$ | Argon | O | 5.5445(1) | 7.7574(2) | 5.5864(2) | 60.051(1) | 0.01 | 112 |
| $La_{0.95}Na_{0.05}Mn_{0.95}Ru_{0.05}O_3$ | Argon | O | 5.5153(2) | 7.8082(2) | 5.5154(2) | 59.384(1) | 0.072 | 182 |
| $La_{0.95}Na_{0.05}Mn_{0.85}Ru_{0.15}O_3$ | Argon | O | 5.5572(1) | 7.7786(3) | 5.5236(2) | 59.753(2) | 0.035 | 167 |
| $La_{0.85}Na_{0.15}MnO_3$ | Oxygen | R | 5.5044(1) | 5.5044(1) | 13.3317(2) | 58.292(2) | 0 | 305 |
| $La_{0.85}Na_{0.15}Mn_{0.95}Ru_{0.05}O_3$ | Oxygen | R | 5.5123(2) | 5.5123(2) | 13.3392(2) | 58.487(1) | -0.02 | 301 |
| $La_{0.85}Na_{0.15}Mn_{0.85}Ru_{0.15}O_3$ | Oxygen | R | 5.5187(1) | 5.5187(1) | 13.3454(2) | 58.645(2) | 0.01 | 172 |
| $La_{0.85}Na_{0.15}MnO_3$ | Argon | R | 5.5094(3) | 5.5094(3) | 13.3363(2) | 58.417(1) | -0.01 | 265 |
| $La_{0.85}Na_{0.15}Mn_{0.95}Ru_{0.05}O_3$ | Argon | R | 5.5179(2) | 5.5179(2) | 13.3461(2) | 58.644(2) | 0.05 | 266 |
| $La_{0.85}Na_{0.15}Mn_{0.85}Ru_{0.15}O_3$ | Argon | O | 5.5246(2) | 7.8091(3) | 5.52191(2) | 59.547(2) | 0.035 | 189 |

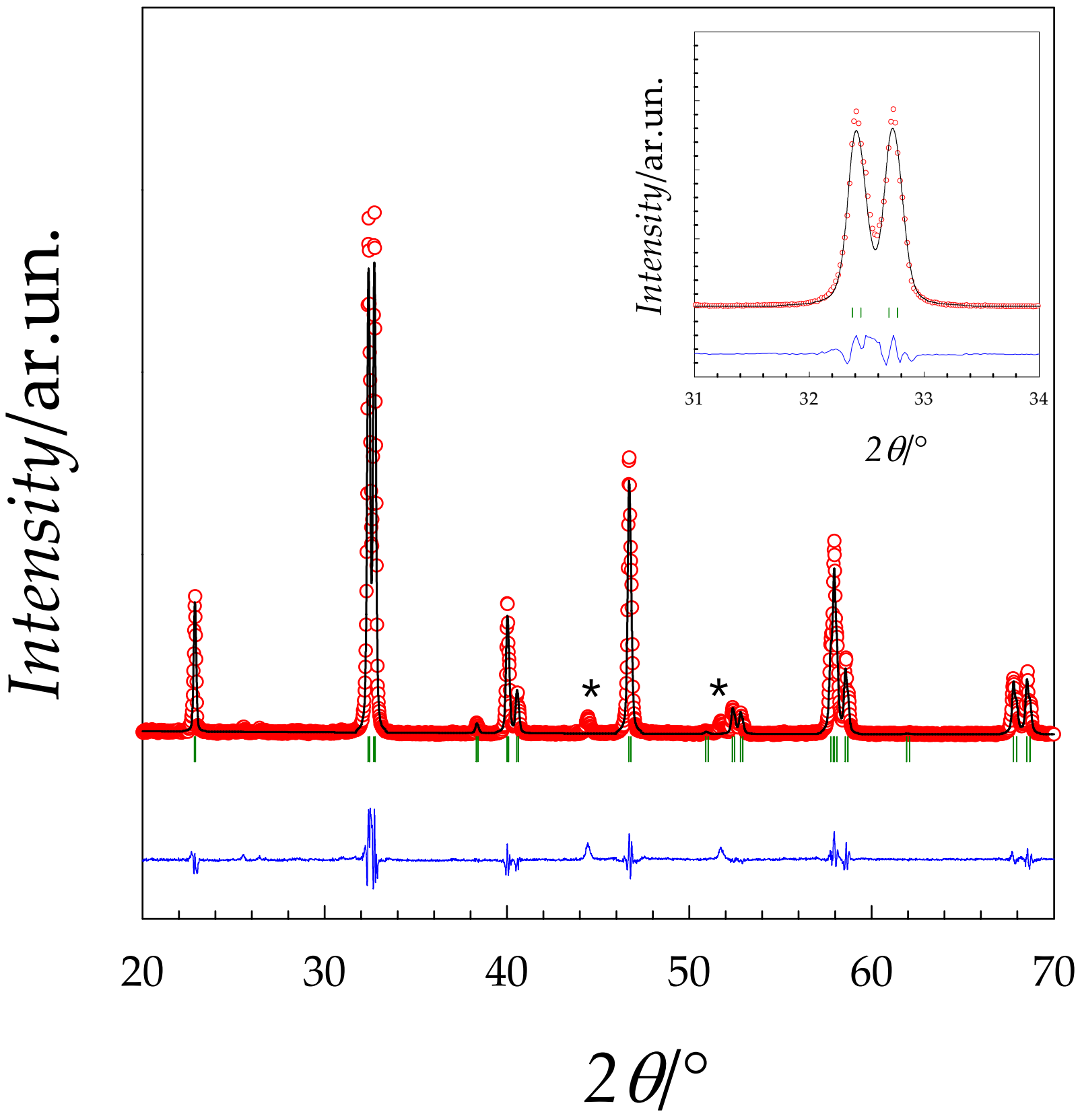

Figure 1

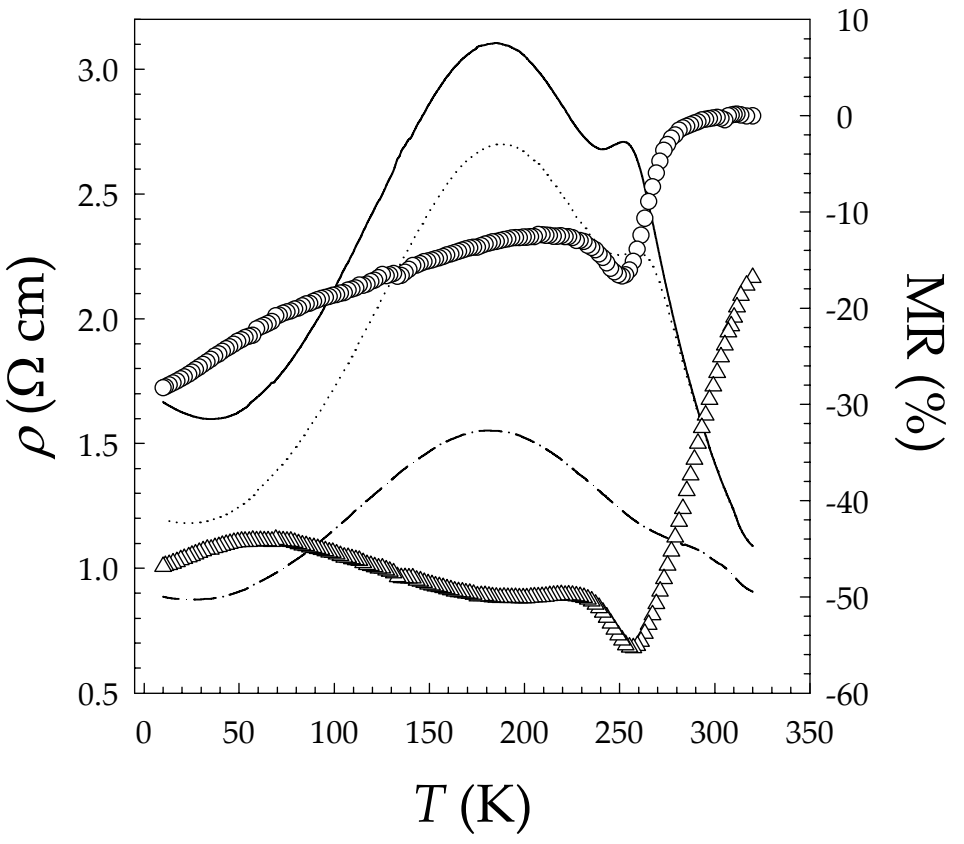

Figure 2a

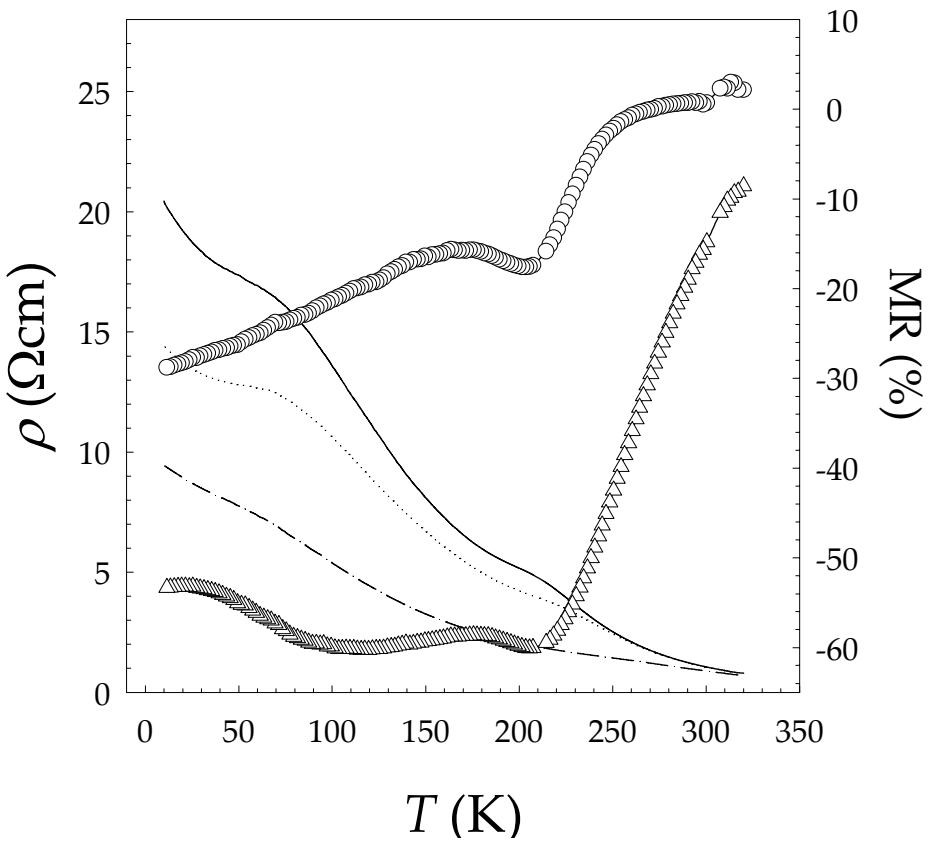

Figure 2b

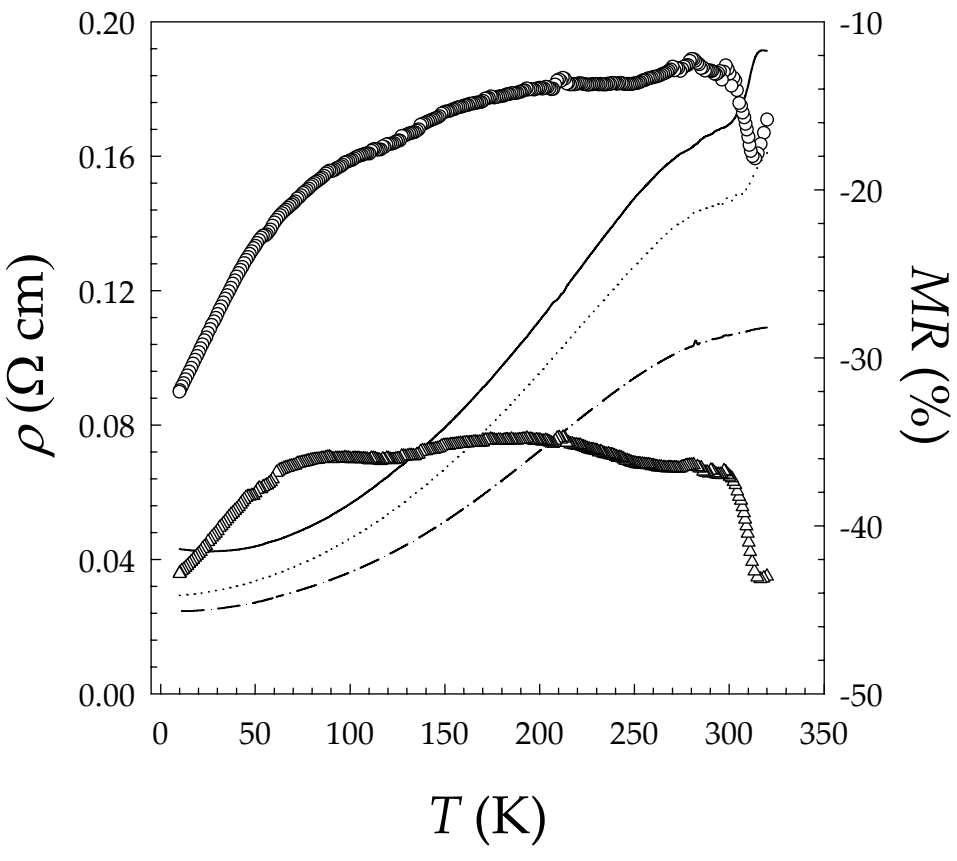

Figure 3a

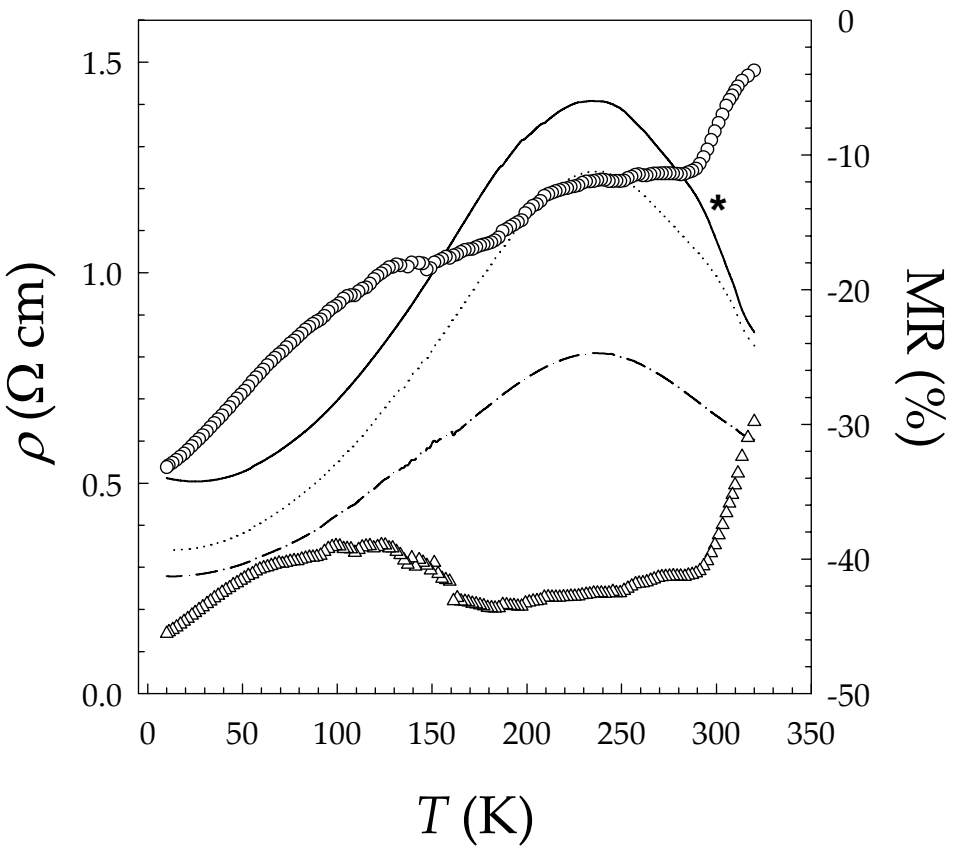

Figure 3b

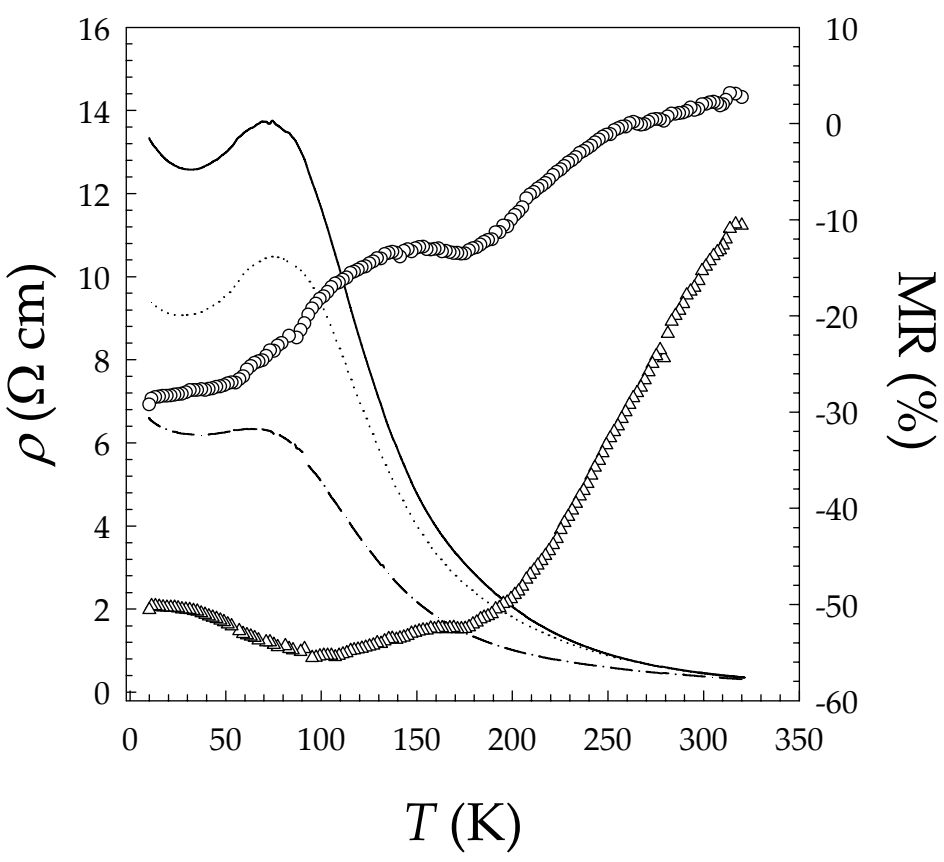

Figure 4a

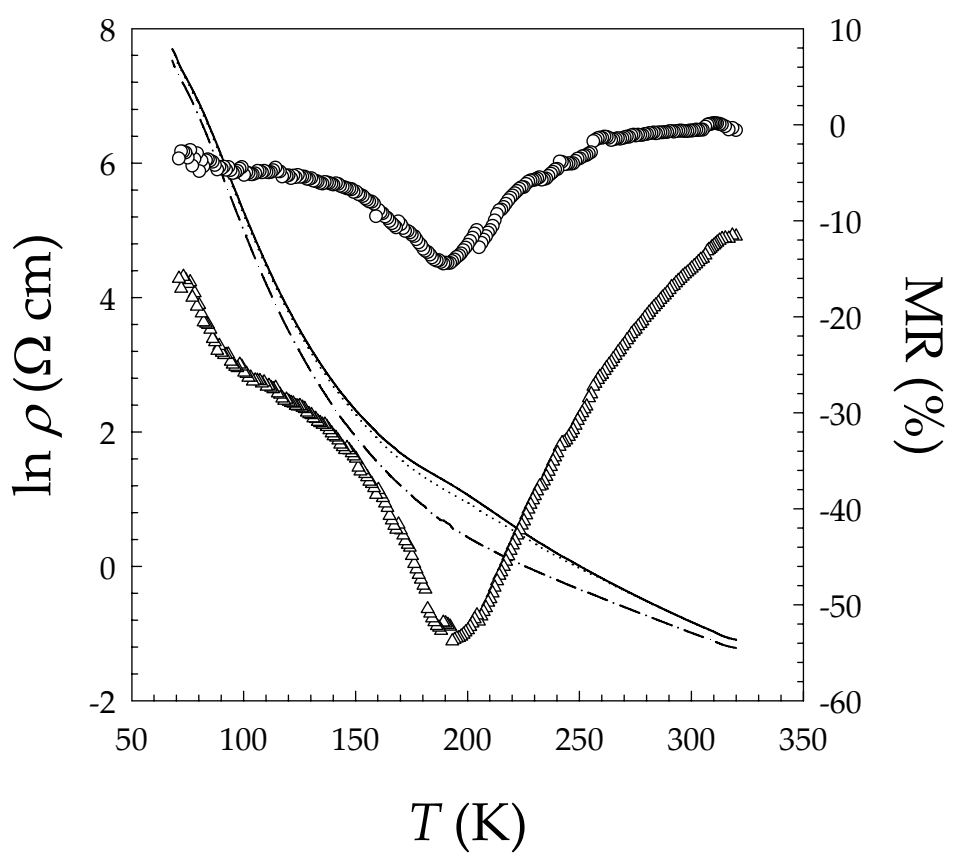

Figure 4b

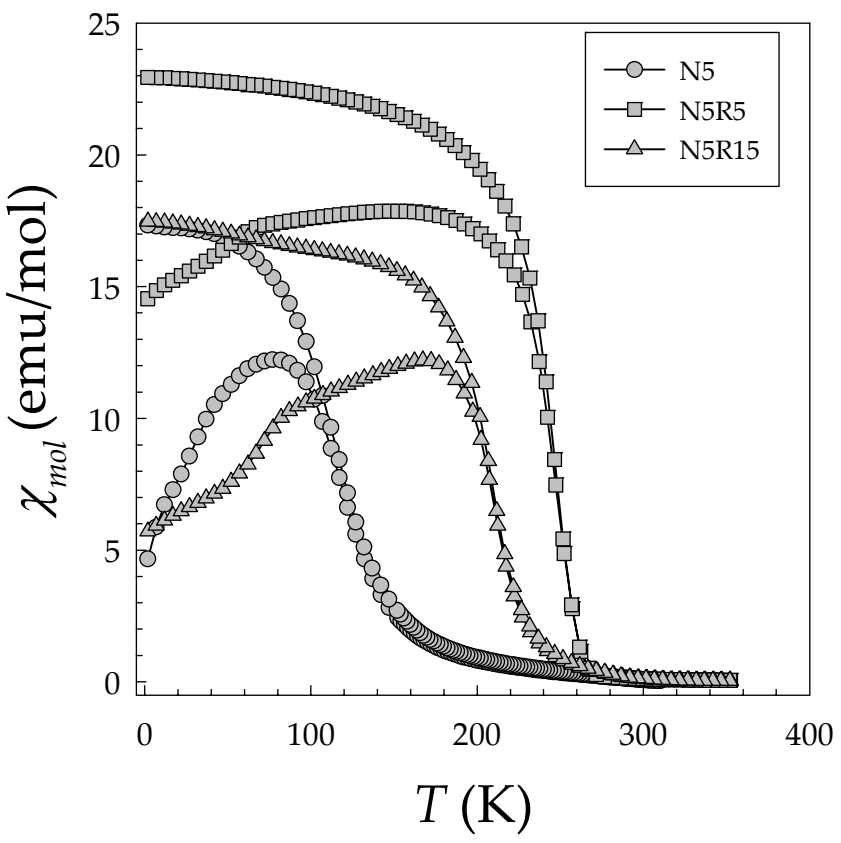

Figure 5a

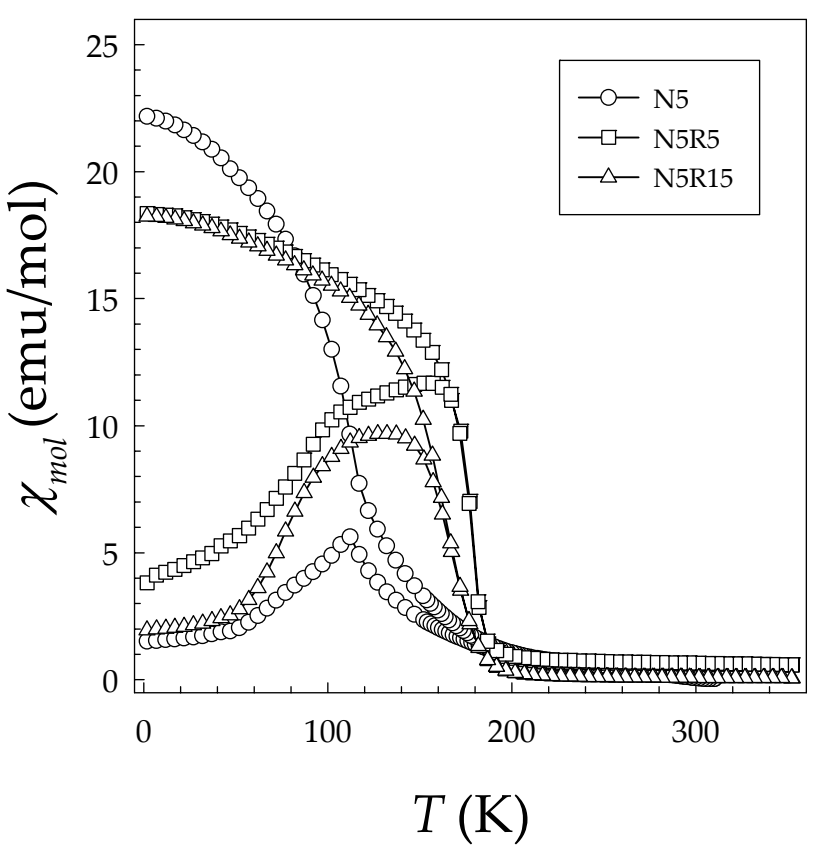

Figure 5b

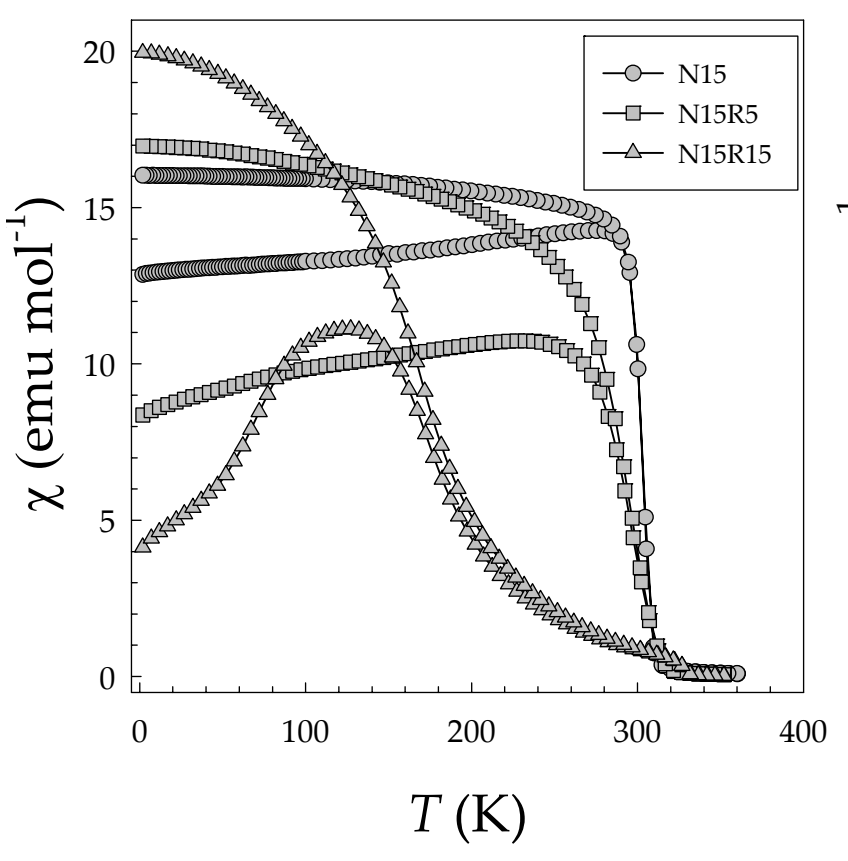

Figure 6a

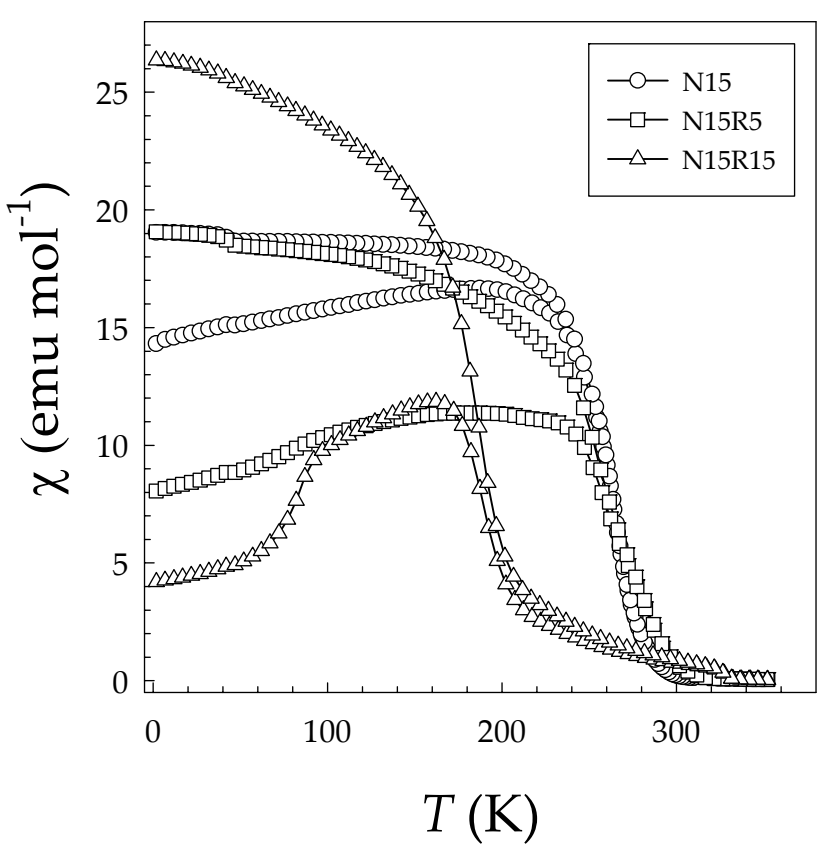

Figure 6b

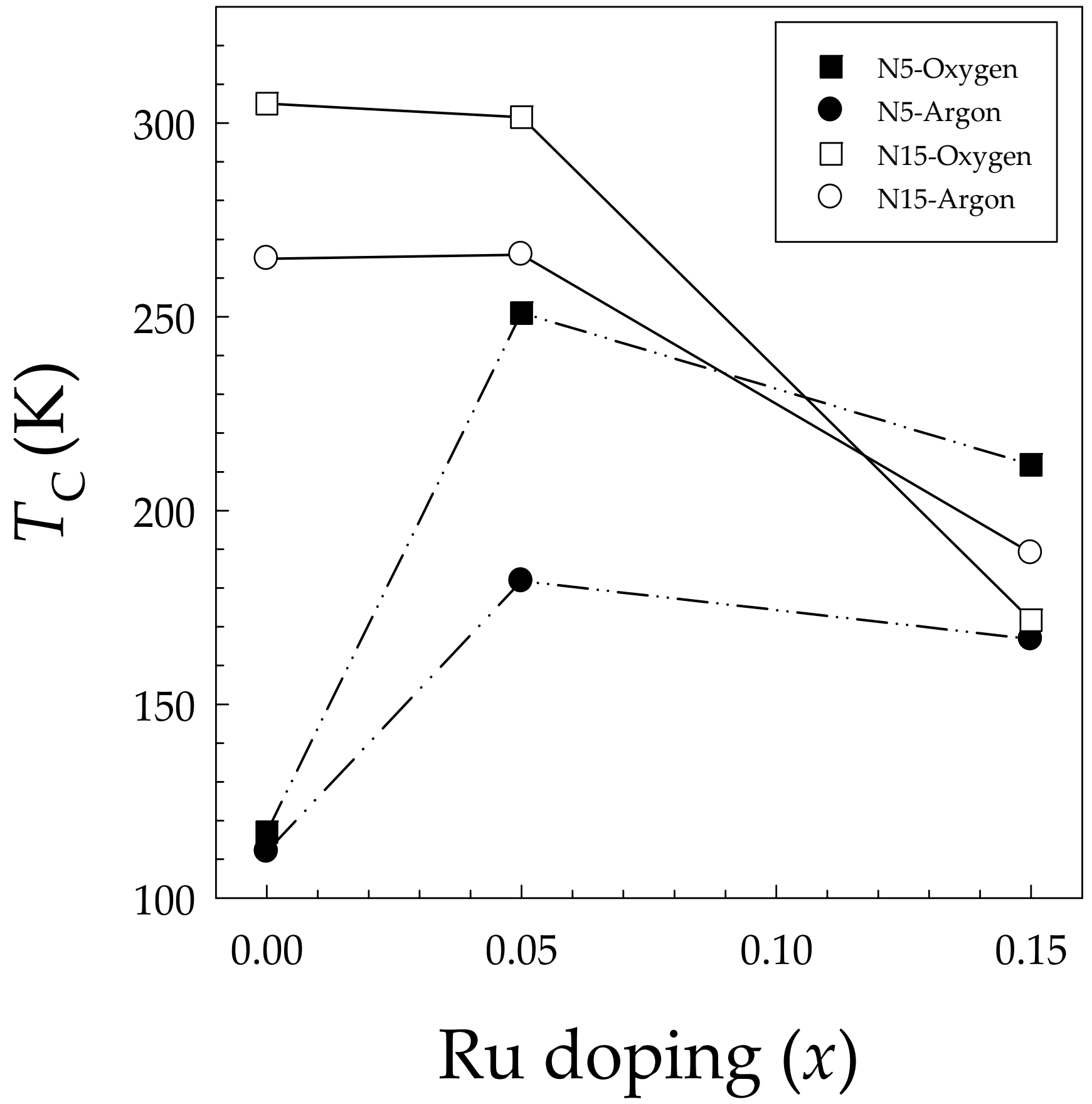

Figure 7